\documentclass[sigconf,nonacm]{acmart}

\usepackage{dsfont}
\usepackage{booktabs}
\usepackage{amsmath}
\usepackage{subcaption}
\usepackage{algpseudocode}
\usepackage{algorithmicx}
\usepackage{algorithm}
\usepackage{multirow}
\usepackage{graphicx}
\usepackage{makecell}
\usepackage[normalem]{ulem}

\begin{document}

\pagenumbering{arabic}

\title{Characterizing Nodes and Edges in Dynamic Attributed Networks: A Social-based Approach}

\author{Thiago H. P. Silva, Alberto H. F. Laender, Pedro O. S. Vaz de Melo}
\email{{thps,laender,olmo}@dcc.ufmg.br}
\affiliation{
  \institution{Department of Computer Science \\ Universidade Federal de Minas Gerais}
  \city{31270-901 - Belo Horizonte - MG} \\
  \country{Brazil} 
}

\renewcommand{\shortauthors}{Silva et al.}

\begin{abstract}
How to characterize nodes and edges in dynamic attributed networks based on social aspects? We address this problem by exploring the strength of the ties between actors and their associated attributes over time, thus capturing the social roles of the actors and the meaning of their dynamic interactions in different social network scenarios. For this, we apply social concepts to promote a better understanding of the underlying complexity that involves actors and their social motivations. More specifically, we explore the notion of \textit{social capital} given by the strategic positioning of a particular actor in a social structure by means of the concepts of \textit{brokerage}, the ability of creating bridges with diversified patterns, and \textit{closure}, the ability of aggregating nodes with similar patterns. As a result, we unveil the differences of social interactions in distinct academic coauthorship networks and questions \& answers communities. We also statistically validate our social definitions considering the importance of the nodes and edges in a social structure by means of network properties.
\end{abstract}


\maketitle
\settopmatter{printfolios=true}

\section{Introduction}

The large amount of data available today from Internet services and applications has allowed us to explore how entities relate to each other. As such, we can map entities and their links as social networks in order to provide new kinds of analysis, both structural and social~\cite{Easley2010}, for instance, to characterize entities behaving like \textit{hubs}~\cite{Newman2004} or acting as \textit{bridges} by connecting different parts of a network~\cite{granovetter1973strength}. Indeed, the so-called social network analysis has contributed to understand how highly connected complex networks work, ranging from graph theory to property rights \cite{watts1998collective,newman2006structure,thesisThiagoSIlva}. In this regard, based on how entities play structural roles in networks, we contribute with a social-based perspective in order to better analyze the behavior and the strength of the ties involving such entities.

We recall that in general social networks are constructed considering the existence of explicit relationships (e.g., social ties with relatives). In this static structural scenario, one approach is to explore the notion of social capital given by the  position of the nodes in the social network structure~\cite{granovetter1973strength,coleman1988social,burt1992structural}. For instance, Granovetter~\cite{granovetter1973strength} defines the concept of \textit{weak ties} as being those important relationships that make a network more cohesive by means of the creation of bridges between communities. As discussed by Aral~\cite{aral2016future}, the most influential sociological theories explore bridging ties (e.g., connecting different parts of a network) and cohesive ones (e.g., building a trust circle), which provide more advantage when accessing information passing through a network.

However, as explicit relationships evolve to other kinds of interactions (e.g., encounters, phone calls, exchanged messages, etc.), they become more complex, thus bringing more information about these social interactions. In this way, a more general approach is required to model these specificities by using only edge or node attributes~\cite{aggarwal2016edge,Rezaei2017}. By doing so, it is possible to promote more information about the social motivation involving each interaction, since individuals tend to change their attributes over time, whether in terms of location (a new job or country), relationships with other people (childhood friends who no longer participate in their network) or new skills acquired. This would enable us to understand the evolution of social structures, in which the persistence of attributes over time indicates the social value associated with each interaction.

Additionally, several other works have investigated topological properties and patterns of social networks in order to define the behavior of their actors and measure the strength of their relationships~\cite{Newman2004,barabasi2009scale,huang2018Triadic,LeaoJISA2018,levchuk2012learning}. Exploring the behavior and the dynamics of the actors in a social network is essential for a better understanding of its social structure, which is usually characterized by graphs that capture the social aspects involved~\cite{medo2016identification,silva2015authorship,YangX16}. Accordingly, Barabási~\cite{barabasi2009scale} reinforces the importance of the network theory paradigm  as fundamental for understanding the complexity that involves actors and their relationships. For example, Newman~\cite{Newman2004} measures the influence of the nodes in a network based on their proximity and the number of shortest paths among them.

In a previous work~\cite{SilvaASONAM2019,silva2020knowledge}, we analyzed how social aspects impact knowledge transfer in a network. Our proposed model allows one to represent the social dynamics of node-attribute relationships to capture the influence generated by knowledge transfer. Here, we go a step forward and focus on the strength of nodes and their dynamic relationships over time by means of social capital.

More specifically, in this article, we propose to address together distinct aspects: actors, interactions, time intervals, attributes and social concepts. Our strategy to address theses issues consists in modeling nodes by associating them with their attributes in order to extract persistent features over time. Regarding its social perspective, our method is based on Burt's definition of social capital by considering two concepts: \textit{closure}, the ability of aggregating actors with similar patterns, and \textit{brokerage}, the ability of creating bridges with diversified patterns~\cite{burt2005brokerage}.

\begin{figure*}[!t]
    \centering
    \includegraphics[width=0.67\textwidth]{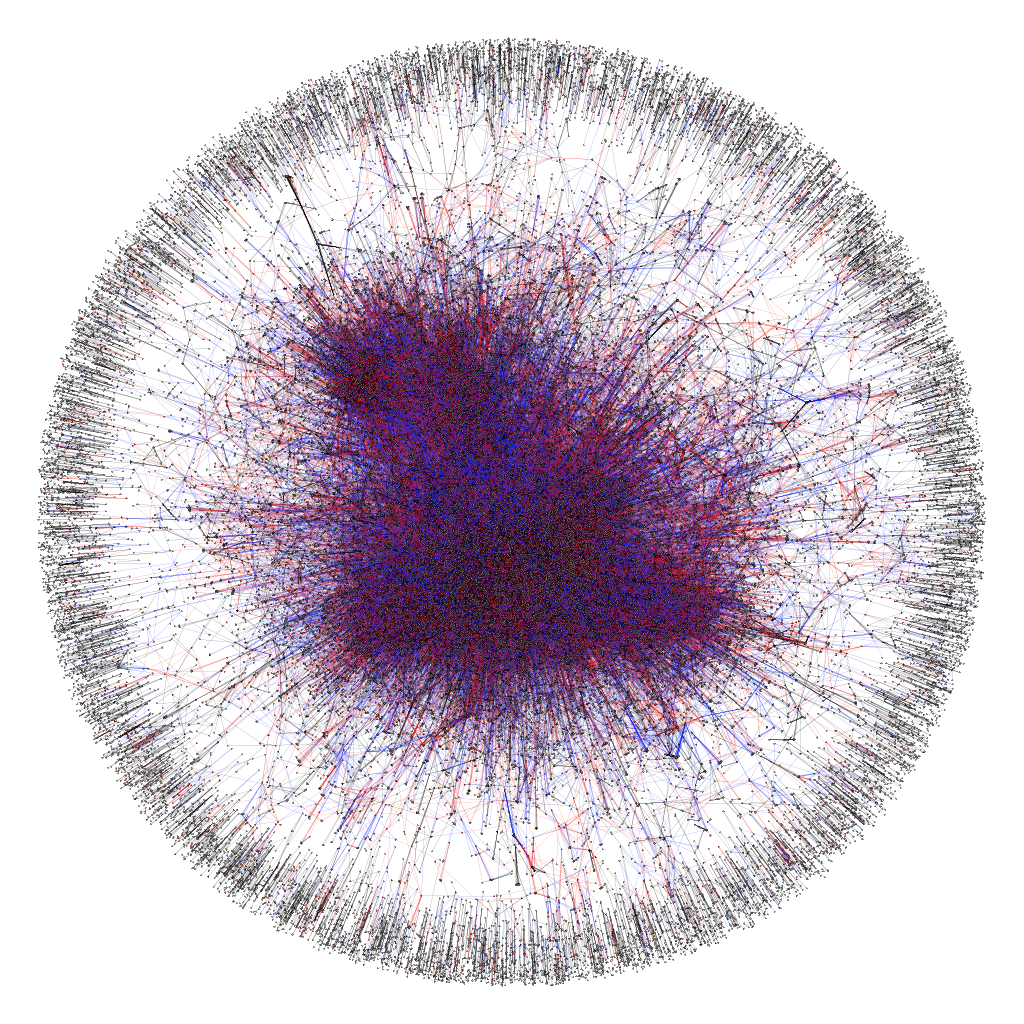}
        \caption{Visualization of a Computer Science academic coauthorship network according to our  proposed social classification. Blue edges emphasize the \textit{closure} concept (strong ties) and red  ones the \textit{brokerage} concept (weak ties). Black edges correspond to those regarded as  \textit{innocuous} (i.e., edges that have no important information passing through them). We have omitted multiple edges connecting the nodes for a better visualization. \textbf{Best viewed in color.}}
        \label{fig:undirectedGraph}
\end{figure*}

In a preliminary work \cite{silvaBigData2018}, we presented two specific contributions to this kind of analysis: (i)~a node-attribute graph model that captures the social tie of individuals and their associated attributes, thus providing a dynamic attributed model that enables us to mine multiple interactions over time; and (ii)~a new method to classify nodes and their relationships based on temporal node-attributes that considers the social role of the nodes and the social meaning of the edges.

Here, we extend our previous work by first redefining the classes assigned to nodes and edges in order to reflect the social concepts behind the proposed model. For this reason, both nodes and edges are now thoroughly classified according to the concepts of \textit{closure} and \textit{brokerage}. We also contribute by applying our new social classification strategy to different types of social network, therefore providing a detailed discussion underlying social aspects based on additional experiments.

In order to illustrate this, Figure~\ref{fig:undirectedGraph} shows a Computer Science academic coauthorship network, including 
more than 79 thousand nodes and 263 thousand edges, built on data collected from DBLP (details presented in Section~\ref{sec:datasets}). In this network, the edges classified according to the concept of \textit{closure} are shown in blue, those classified according to the concept of \textit{brokerage} are shown in red and those that express no social meaning (i.e., that are non-relevant) are shown in black. Note that the edges based on the \textit{brokerage} and \textit{closure} concepts dominate the center of the graph, while the extremities tend to have a greater prominence of edges regarded as non-relevant. This means that edges strongly related to social concepts tend to be better positioned in a social structure (i.e., linked to central nodes), which provides early access to information passing through the network. Our approach allows to analyze networks in terms of their \textit{structural autonomy}, which occurs when people are tightly connected to one another with extensive bridge ties beyond them, thus achieving high levels of innovation and productivity because there are both trust and cooperation between individuals who connect different parts of a network~\cite{burt2005brokerage}.

The rest of this article is organized as follows. Firstly, Section~\ref{sec:related} reviews related work, whereas Section~\ref{sec:basicdefinitions} introduces our proposed model and describes the methodology adopted for evaluating it. Then, Section~\ref{sec:analysis} analyzes and discusses the results of applying our classification method to different social network scenarios, whereas Section~\ref{sec:experiments} summarizes the results of our experimental validation. Finally, Section~\ref{sec:conclusions} presents our conclusions and some considerations for future work.

\section{Related Work}\label{sec:related}

The study of the individuals' dynamics enables us to understand the evolution of social networks, thus providing models to characterize their behavior~\cite{thesisThiagoSIlva,iacobelli2017edge,medo2016identification,silva2015authorship,barabasi2002evolution}. For instance, Barab{\'a}si et al.~\cite{barabasi2002evolution} capture the social tie importance by observing the topology and the nodes' internal behavior in coauthorship networks, whereas Sun et al.~\cite{sun2013social} propose a model to analyze the social dynamics of science in terms of scientific disciplines. Yet, Iacobelli and Figueiredo~\cite{iacobelli2017edge} explore random walks on dynamic networks to better characterize and understand their structures. In another context, Silva et al.~\cite{silva2015authorship} characterize the moving properties and the behavioral profile of how researchers move around publication venues stratified in terms of their quality, whereas Brandão et al.~\cite{brandao2017tie} address how the social roles of researchers change over time. Also, traditional network metrics have been employed to identify the most important nodes within a graph, such as done by Newman~\cite{Newman2004} that uses centrality metrics based on shortest paths (e.g., closeness and betweenness) for determining the best positioned nodes in academic social networks.

Several other studies analyze social networks based on particular social concepts such as tie strength~\cite{brandao2017tie,brandao2015analyzing,gilbert2009predicting,RECAST2013}, the friendship paradox~\cite{adamic2003friends,higham2019centrality} and social influence~\cite{JiangSAYW17}, as well as on structures such as triadic closure and social balance~\cite{huang2018Triadic,giscard2017evaluating}. For instance, based on the information shared between nodes, Adamic and Adar~\cite{adamic2003friends} measure the strength of relationships by analyzing the similarity between messages exchanged between individuals. Yet, Gilbert and Karahalios~\cite{gilbert2009predicting} also consider temporal aspects, but modelling tie strength as a linear combination of different dimensions such as intensity by means of the number of words exchanged, emotional support based on positive words, social distance in terms of the political differences, among other aspects. Finally, Levchuk et al.~\cite{levchuk2012learning} propose an approach to learn and detect network patterns such as repetitive groups of people involved in coordinated activities.

Regarding social perspectives, several works have explored the notion of social capital given by the strategic positioning of a particular actor in a social structure~\cite{granovetter1973strength,feng2018identification,burt2005brokerage}. Based on the premise that actors can make a network stronger by integrating different parts, Granovetter~\cite{granovetter1973strength} defines the concept of \textit{weak ties} as being those important relationships that make a network more cohesive by creating bridges between communities. Likewise, Burt~\cite{burt2005brokerage} describes a \textit{structural hole} as the gap formed by individuals who have complementary knowledge, and then defines as \textit{brokers} those nodes that hold certain positional advantages due to their good location in the social structure. Considering such results, Feng et al.~\cite{feng2018identification}, for instance, used structural holes to identify the most central and bridging group of nodes in a network. 

In another context, Brandão et al.~\cite{brandao2017tie} investigate the strength of dynamic social relationships in academic social networks based on topological metrics, thus revealing that such relationships tend to have more weak and random ties than strong and bridges ones. Leão et al.~\cite{LeaoJISA2018} analyze the role of random interactions in the structure of communities, whereas Sanz-Cruzado and Castells~\cite{sanz2018enhancing} analyze the role played by strong ties (links within communities) and weak ties (links between communities), thus showing that bridges work as enhancers of the structural diversity in social networks.

In this article, we define social classes to better characterize social networks. In this regard, Yang et al.~\cite{yang2014sparrows} propose a metric for expert identification in the StackOverflow Q\&A site, which is based on the quality of its users' contribution. Specifically, they define two profiles: \textit{sparrows}, as being highly active users on the network that contribute to the vast majority of its content, and \textit{owls}, as being the most experienced users that provide useful answers. Furtado et al.~\cite{FurtadoCSCW13} also characterize users' behavior in Q\&A sites, but observing a dynamic perspective. They define ten profiles based on motivation (e.g., activity duration) and ability (e.g., how useful a comment is) metrics, thus characterizing users as unskilled, expert, activist and hyperactivist. In another context, Vaz de Melo et al.~\cite{RECAST2013} propose the RECAST (\textit{Random rElationship ClASsifier sTrategy}) algorithm that identifies
random and social interactions based on network properties. Specifically, such algorithm explores topological and temporal
aspects in order to measure the strength of the nodes’ relations, which
is derived from the neighborhood overlap
and the persistence of the relationships. 
Doing so, it is able to classify the edges of a network by assigning them to one of the following social classes: \textit{friend}, \textit{acquaintant}, \textit{bridge} and \textit{random}.

Our proposed method also relies on temporal aspects and on the regularity of the relationships over time, but differs from those discussed above by mainly exploring social concepts. In this regard, Alhazmi and Gokhale~\cite{alhazmi2016mining} measure the structural social capital of online social networks by defining closure and brokerage as closed and open triads, respectively. Alternatively, in a recent work~\cite{SilvaASONAM2019}, we classify nodes and edges by inspecting how knowledge is transferred across a network. For this, we depict a closure tie when two individuals are teaching to and learning from each other, whereas a brokerage tie characterizes a knowledge transfer between an expert and an inexpert user. Here, instead of describing an entire network or exploring knowledge-transfer behaviors, we aim at characterizing nodes and edges based on the strength of the social ties with their relevant attributes. Thus, we explore strong ties between nodes and their relevant attributes as representing the closure effect, and weak ties underlying the potential of the knowledge acquired by the nodes in the network as depicting the brokerage effect.

\section{Social-based Perspective}\label{sec:basicdefinitions}

In this section, we first present an overview of our method for social-based classification of multiple interactions in dynamic attributed networks~\cite{thesisThiagoSIlva,silvaBigData2018}. Then, we present and discuss our classification scheme entirely based on social concepts. 
For this, we reinforced the notion of social capital to capture the importance of the nodes based on their positioning in the network structure and the social meaning of the relationships involved. 

\subsection{Modeling Dynamic Interactions} 

In our previous work~\cite{silvaBigData2018,silva2020knowledge}, we modeled attributed networks by a dynamic multigraph $G = \bigcup_{1}^{t}\mathcal{G}_i$ that represents a temporal aggregated graph within a time interval~$[1,t]$. In short, $G$ is the union of temporally disjoint graph instances $\mathcal{G}_k = (\mathcal{V}_k, \mathcal{E}_k)$ constructed in time steps of size $k$ (e.g., days)~\cite{RECAST2013}. Likewise, in order to consider the dynamic temporal attributes, we defined a heterogeneous dynamic multigraph graph $H = \bigcup_{1}^{t}\mathcal{H}_i$, where $\mathcal{H}_k = (\mathcal{V'}_k, \mathcal{E'}_k)$. This graph consists of two types of node: actors (e.g., researchers) and attributes (e.g., expertise). In other words, this strategy is an abstraction that transforms the attributes of each edge into additional nodes, allowing an original actor node to be directly connected to these new attribute nodes. In this context, each actor is associated with a set of attributes that can change overtime. We also assume that all attributes are related to some knowledge, which can be seen as a skill acquired by the actors along the time. 

\subsection{Extracting Relevant Attributes} 

The next step in our approach is to determine the set of relevant attributes for each node at each time interval. We define as relevant attributes those that are closely connected to the nodes, i.e., persistent in their histories. The idea is to identify, for each actor, all attributes and evaluate them in terms of the set of attributes most strongly statistically associated with the actor nodes (i.e., according to their stability along the time).

For this, we analyze the nodes' interaction history in order to extract knowledge from the node-attribute relationships. We apply the concept of persistence of an edge along the time, which captures the importance of the relationship between two nodes in terms of their associated attributes. The \textit{persistence} metric of an edge is defined as $pers_t(u,a) = \frac{1}{t}\sum_{k=1}^{t} \mathds{1}_{\mathcal{E}^{'}_k}((u,a))$, where the indicator function is defined as
\begin{equation}
\mathds{1}_{\mathcal{E}^{'}_k}((u,a)) =
\begin{cases}
1, \text{if } (u,a) \in \mathcal{E}^{'}_k,\\
0, \text{otherwise.}\\
\end{cases}
\end{equation}

\noindent Note that this operation is performed on each attributed graph at discrete intervals and not on the aggregated graph. In other words, it captures the dynamics by observing the persistence in each temporal subgraph within the time interval $[1,t]$.

\algdef{SE}[SUBALG]{Indent}{EndIndent}{}{\algorithmicend\ }%
\algtext*{Indent}
\algtext*{EndIndent}
\begin{algorithm}[t]
\caption{Extracting Relevant Attributes}
\label{algo:extrairAtributos}
\begin{algorithmic}[1]
\algtext*{EndIf}
\algtext*{EndFor}
\Require $H$, $t$
\Ensure $\Gamma_k(u)$, $\forall u \in \bigcup_{k=0}^{t}V_k$
\ForAll {$u \in V_t$}
    \State $\mathcal{A}_{temp} \gets \{\}$
    \ForAll {$k \in [1, t]$}
    	\State $\Gamma_k(u) \gets \{\}$
    	\State $\mathcal{A}_{temp} \gets \mathcal{A}_{temp} \cup \{ a | (u, a) \in \mathcal{E}^{'}_k \}$
    	\State $vector \gets \{\}$
        \ForAll {$a \in \mathcal{A}_{temp}$}
        	\State $vector.add(pers_k(u, a))$
        \EndFor
        \State $IQR \gets percentile(vector, 75) - percentile(vector, 25)$
        \ForAll {$a \in \mathcal{A}_{temp}$}
            \If {$pers_k(u, a) > percentile(vector, 75) + IQR*1.5$}
        		\State $\Gamma_k(u) \gets \Gamma_k(u) \cup \{a\} $
            \EndIf
        \EndFor
    \EndFor
\EndFor
\end{algorithmic}
\end{algorithm}

More precisely, Algorithm~\ref{algo:extrairAtributos} details the process of extracting relevant attributes. It receives as input the aggregated graph $H = \{\mathcal{H}_1, ..., \mathcal{H}_t\}$ and the final time interval $t$. In summary, the algorithm inspects, for each actor, all attributes and evaluates them according to their persistence along the time by means of percentiles (function \textit{percentile} on lines 9 and 11), thus identifying the set of attributes most strongly statistically associated with the actor's nodes. The idea is to filter such attributes that are exaggeratedly linked to a node in a specific period in comparison to the others, i.e., identifying the abnormal presence of certain attributes at each time point. In order to choose the appropriate statistical method to select the most significant attributes, we first check whether the values of the edge persistence metric follows a normal distribution. Then, we extract the relevant attributes based on the definition of an outlier given by the interquartile range (IQR). Another approach is to use the modified z-score for the same purpose~\cite{iglewicz1993detect}. Since the experimental results were similar for IQR and for the modified z-score, we chose IQR due to the possibility of applying different percentages by means of percentiles (i.e., adapting the constraints according to specifics problems).

As a result, this strategy builds a set comprising all attributes statistically relevant for each node $u \in G$ at a time interval $k$ (i.e., for each subgraph), which are referenced as $\Gamma_k(u)$. Note that the sets $(\Gamma_1(u), \Gamma_2(u),..., \Gamma_t(u))$ are dynamically built according to the degree of persistence, i.e., different instants $k$ may contain completely distinct sets of attributes.

\noindent
\subsection{The Classifier} Our classification scheme reinforces the importance of social concepts as a relevant factor for better understanding the complexity that involves actors and their relationships in dynamic attributed networks. In summary, we classify nodes and edges as follows:

\begin{itemize}    
    \item \textit{Node classification.} This classification captures the expertise of the nodes by means of their relevant attributes. For example, in an academic social network, a node that has a long-lasting association with attributes like \textit{relational model}, \textit{data definition} and \textit{query language} is likely to have an authority over them. Thus, this node can be classified as having a \textit{strong tie} with the \textit{Databases} domain.
    
    \item \textit{Edge classification.} This classification assigns a tie strength to the edges in order to represent the kind of their interactions. For example, in a social media network like Facebook, a \textit{strong} edge may indicate a social tie between relatives, whereas a \textit{weak} one may represent a social tie with a co-worker.
\end{itemize}

Based on the social structure that models the dynamic interactions along the time, the edge classification process assesses the degree of relevance of the attributes associated with each node by considering its past interactions (see Algorithm~\ref{algo:extrairAtributos}). For this, it determines the dynamic state of a node at each time interval as representing a \textit{strong}, \textit{weak} or \textit{non-relevant} association with a specific attribute (knowledge). In this context, the \textit{strong} state represents the importance of a node in terms of its expertise within a closed group, whereas the \textit{weak} one captures its potential for connecting different parts of a network. The next step consists in mapping these dynamic states in order to determine the social classes of the edges. We propose three social classes for edges\footnote{In our previous work~\cite{silvaBigData2018}, we defined seven classes for the edges (\textit{very strong}, \textit{strong}, \textit{strong bridge}, \textit{regular bridge}, \textit{weak bridge}, \textit{ordinary} and \textit{sporadic}), but we report that they were not very discriminatory. Thus, here we map the classes according to Burt's social theory~\cite{burt2005brokerage}, which provided more representative results.}: \textit{closure}, \textit{brokerage} and \textit{innocuous}. In this way, such social classes emphasize the strength of the relationships as strong ties (\textit{closure}), weak ties (\textit{brokerage}) and non-relevant (\textit{innocuous}), the latter when there is no relevant information passing through the edge.

\begin{algorithm}[t]
\caption{Classifying Edges}
\label{algo:classEdges}
\begin{algorithmic}[1]
\algtext*{EndIf}
\algtext*{EndFor}
\Require $G$, $t$, $\Phi$ e $\Gamma$
\Ensure $\Delta((u, v)), \forall (u, v) \in {\bigcup_{k=1}^{t}\mathcal{E}_k}$
\ForAll {$k \in [1, t]$}
	\ForAll {$(u, v)  \in \mathcal{E}_k $}
		\If {$|\Gamma_k(u)| \neq 0$}
            \State \textbf{if} $|\Gamma_k(u) \cap \Phi((u, v))| \neq 0$
            \Indent 
				\State \textbf{then} $u_{state} \gets strong$
			\EndIndent
            \State \textbf{else} $u_{state} \gets weak$
        \EndIf
        \State \textbf{else} $u_{state} \gets$ \textit{non-relevant}
		\If {$|\Gamma_k(v)| \neq 0$}
			\If {$|\Gamma_k(v) \cap \Phi((u, v))| \neq 0$}
            	\State $v_{state} \gets strong$
            \EndIf
            \State \textbf{else} $v_{state} \gets weak$
        \EndIf
        \State \textbf{else} $v_{state} \gets$ \textit{non-relevant}
		\If {$u_{state} = strong$ \textbf{or} $v_{state} = strong$}
			\State $\Delta((u, v)) \gets closure$
        \ElsIf {$u_{state} = weak$ \textbf{or} $v_{state} = weak$ }
        	\State $\Delta((u, v)) \gets brokerage$
        \EndIf
        \State \textbf{else} $\Delta((u, v)) \gets innocuous$
	\EndFor
\EndFor
\end{algorithmic}
\end{algorithm}

Formally, Algorithm~\ref{algo:classEdges} describes our process for classifying multiple edges. Note that we express the set of attributes of the edge formed by the nodes $u$ and $v$ at time $k$ as the function $\Phi_k((u, v))$. In addition to function $\Phi$, the algorithm receives as input the multigraph $G$, the final time $t$ and the function $\Gamma$ that defines all relevant attributes for each node at each instant $k$ (Algorithm~\ref{algo:extrairAtributos}). First, the algorithm determines the nodes' dynamic states. These states are assigned independently at each iteration of the algorithm and considering each instant $k$ in which an edge is inspected. A node is assigned a state \textit{strong} when there is a strong temporal link with its attributes at the exact moment of the interaction (lines 4 and 5, and 9 and 10). However, if these attributes do not apply to the inspected edge, then the state \textit{weak} is assigned to it (lines 6 and 11). If there are no relevant attributes and the node is active in more than one time interval, then the state \textit{non-relevant} is assigned to it (lines 7 and 12). Once the dynamic states have been assigned to nodes $u$ and $v$, the class of the corresponding edge $e$ is assigned according to them (lines 13-17). More specifically, the \textit{brokerage} class can be seen as a social tie of nodes from distinct domains (lines 13 and 14), whereas the \textit{closure} one establishes a social role by demonstrating a high tightness between a node and its attributes (line 15 and 16). Finally, an \textit{innocuous} class means that there is no knowledge being disseminated through the inspected relationship (line 17).

For classifying the nodes, the same classes are assigned to them, in which case we mean by \textit{closure} a node that has authority on certain attributes, by \textit{brokerage} a node that has a weak association with its attributes and by \textit{innocuous} a node that has an occasional presence in the network. The function~$\Omega$ for this node classification is given by
\begin{equation}
\Omega(u) = \left \{ \begin{array}{ll}
        closure, & \text{if } |\Gamma_t(u)| \neq 0\\
        brokerage, & \text{else if } \sum_{k=1}^{t} \mathds{1}_{\mathcal{V}_k}(u) > 1\\
        innocuous, & \text{otherwise,}
        \end{array} \right.
\end{equation}

\noindent 
where the indicator function is defined as
\begin{equation}
\mathds{1}_{\mathcal{V}_k}(u) =
\begin{cases}
1, \text{if } u \in \mathcal{V}_k,\\
0, \text{otherwise.}\\
\end{cases}
\end{equation}

In summary, the aforementioned social classes reinforce a sociological perspective based on their positioning in a social structure~\cite{granovetter1973strength,burt2005brokerage,guimera2005team}, i.e., by applying social concepts to better understand the strength of the \textit{node-attribute} relationships. More precisely, we rely on Burt's definition of social capital~\cite{burt2005brokerage} by considering the concept of \textit{closure} as representing the importance of a node in terms of its expertise within a closed group, whereas the concept of \textit{brokerage} captures its potential for knowledge transfer. In other words, by a \textit{closure} edge we mean a high tightness between a node and its attributes, whereas a \textit{brokerage} edge can be seen as a social tie of nodes with distinct relevant attributes. Likewise, when classifying a node, the \textit{closure} class is assigned to it when there is a strong tie with some knowledge under its set of relevant attributes and the \textit{brokerage} class when it represents a potential to acquire knowledge from attributes outside its own set of relevant attributes. Indeed, strong ties with certain attributes show an authority on them, whereas weak ties indicates a great potential to diffuse knowledge from its domain. Finally, the \textit{innocuous} class assigned to a node or edge represents no skill acquired by an individual or a lack of relevant information passing through a relationship, respectively.

\section{Characterization of Nodes and Edges}\label{sec:analysis}

In this section, we characterize several social contexts based on our proposed classification method. We begin by introducing the datasets considered. Then, we analyze the overall results of our method when classifying the social behavior of the nodes and the social meaning of their interactions.

\subsection{Datasets}\label{sec:datasets}

\begin{table}[t!]
\centering
\caption{Statistics of the social networks considered.}
\label{tab:stats_networks}
\footnotesize
\begin{tabular}{|l|c|c|c|}\hline
\textbf{Network}                     & 
\textbf{\#snapshots} & 
\textbf{\#nodes} &
\textbf{\#edges}  \\ \hline
\multicolumn{1}{|l|}{24 SIG Academic Networks}          &                                      &                                   &                                            \\
Average                                                 & 34.5                                 & 4.0K                              & 10.9K                                      \\
Median                                                  & 35.0                                 & 2.6K                              & 7.1K                                       \\
Standard Deviation                                      & 9.8                                 & 2.9K                              & 9.2K                                       \\ \hline 
\multicolumn{1}{|l|}{Full Academic Network}             & 55                                   & 79.7K                             & 263.1K                                     \\ \hline \hline
\multicolumn{1}{|l|}{28 Q\&A Communities}               &                                      &                                   &                                            \\
Average                                                 & 2.7M                                 & 4.8K                              & 20.9K                                      \\
Median                                                  & 2.7M                                 & 3.6K                              & 13.8K                                      \\
Standard Deviation                                      & 1.2M                                 & 7.0K                              & 23.6K                                      \\ \hline
\end{tabular}
\end{table}

\begin{table*}[t]
\centering
\footnotesize
\caption{Social classification of nodes and edges for the coauthorship networks.}
\begin{tabular}{|l|l|c|c|c||c|c|c|}
\hline
\multicolumn{2}{|c|}{\multirow{2}{*}{Networks}} & \multicolumn{3}{c||}{Nodes}                                 & \multicolumn{3}{c|}{Edges}                                 \\ \cline{3-8} 
\multicolumn{2}{|c|}{}                          & \textit{closure} & \textit{brokerage} & \textit{innocuous} & \textit{closure} & \textit{brokerage} & \textit{innocuous} \\ \hline
\multirow{3}{*}{24 SIGs} & \textit{Average}	& 27.0\% & 15.2\% & 57.8\% & 29.4\% & 35.3\% & 35.2\% \\ \cline{2-8} 
 & \textit{Median} 				& 27.1\% & 14.9\% & 57.2\% & 30.4\% & 34.8\% & 32.3\% \\ \cline{2-8} 
 & \textit{Std. Dev.} 				&  8.5\% &  2.8\% & 10.8\% &  7.0\% & 10.4\% & 14.6\% \\ \hline \hline
\multicolumn{2}{|l|}{Full Network}	& 18.8\% & 12.2\% & 69.0\% & 31.5\% & 37.1\% & 31.4\% \\ \hline
\end{tabular}\label{tab:ClassPerc}
\end{table*}

\begin{figure*}[!t]
    \centering
    \begin{subfigure}[b]{0.495\textwidth}
        \includegraphics[width=\textwidth]{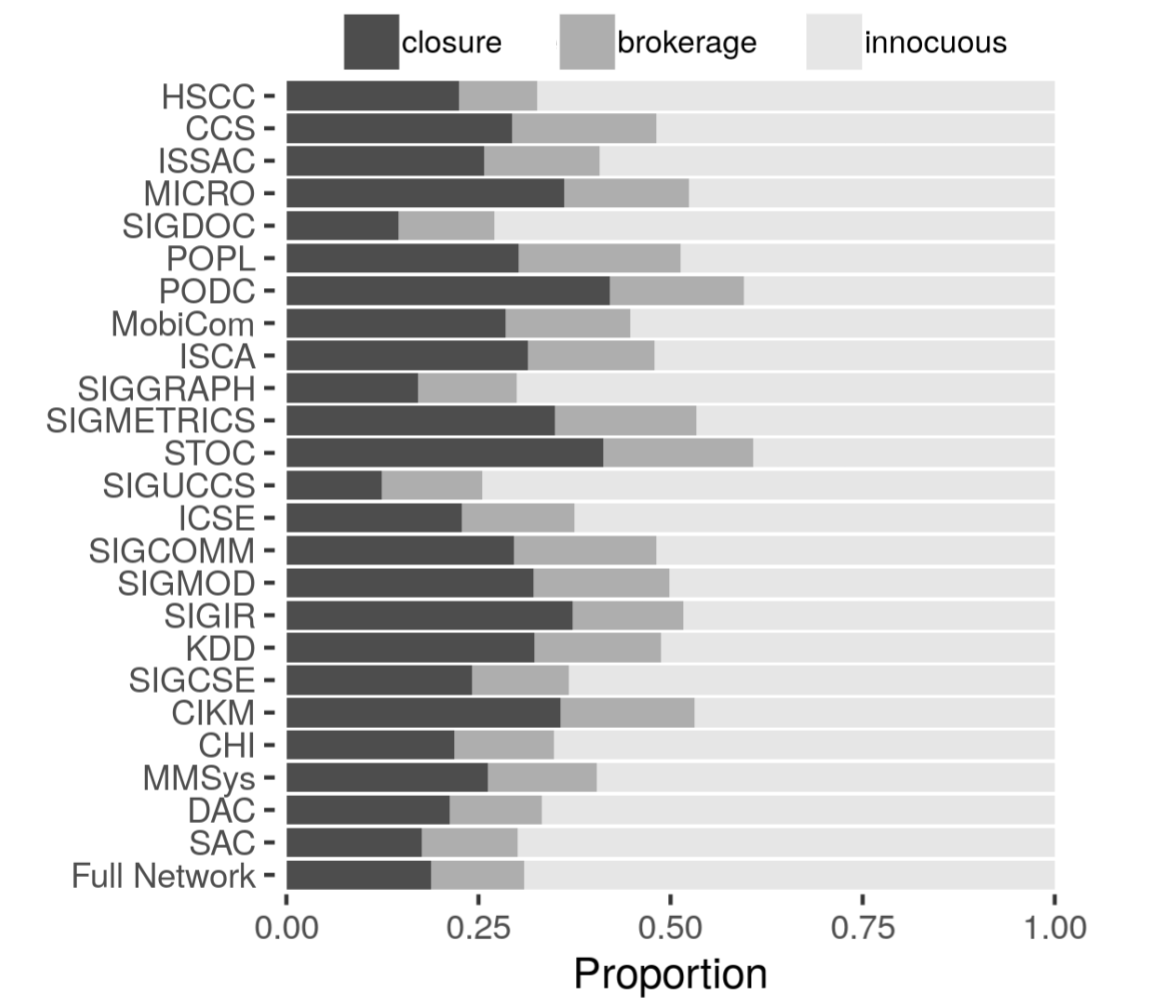}
        \caption{Nodes}
        \label{fig:classNodesAcademic}
    \end{subfigure}
    \begin{subfigure}[b]{0.495\textwidth}
        \includegraphics[width=\textwidth]{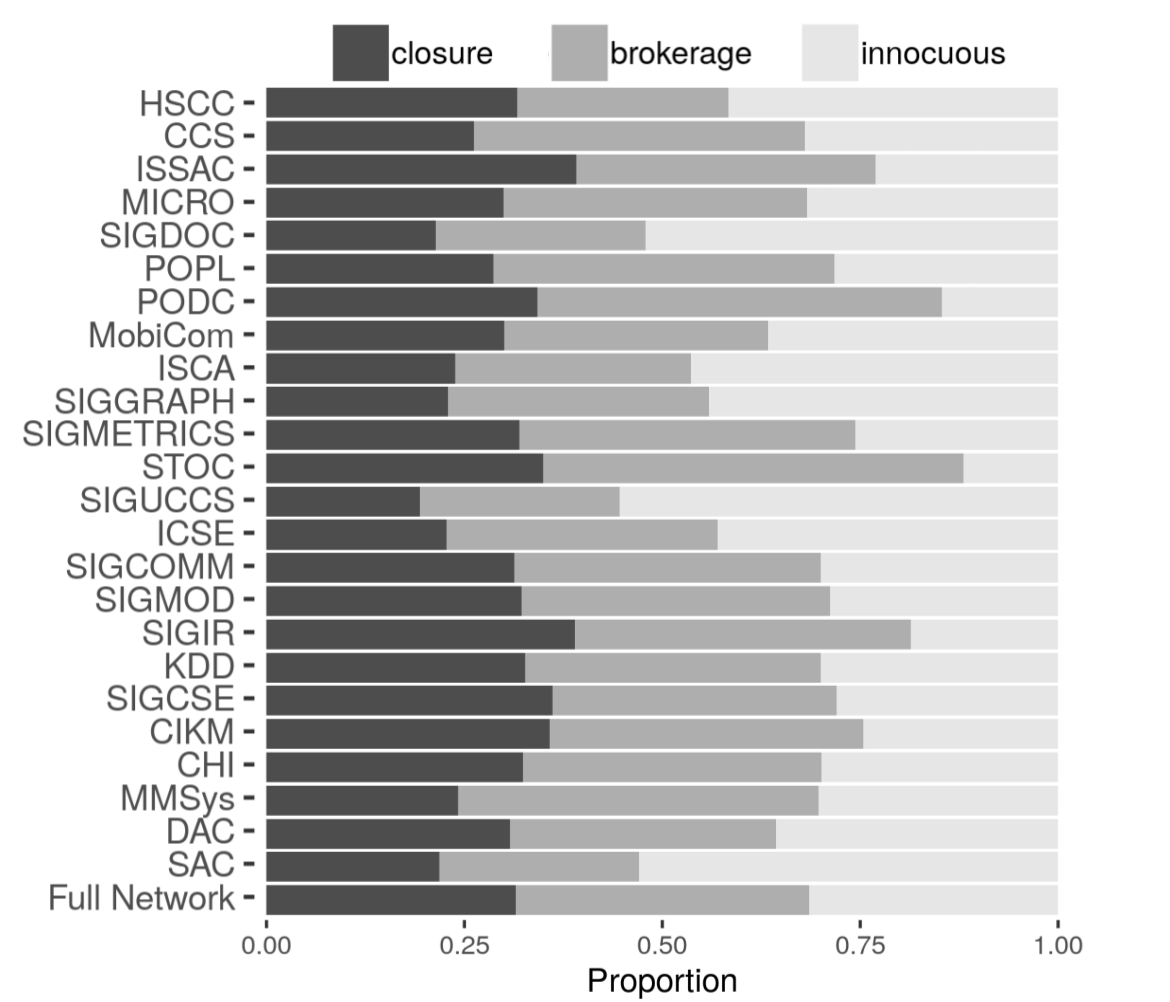}
        \caption{Edges}
        \label{fig:classEdgesAcademic}
    \end{subfigure}
    \caption{Node and Edge classifications for the 24 ACM SIG communities and the full network.}
\end{figure*}

We consider two specific scenarios, academic social networks and Q\&A communities, whose datasets were built in our previous work~\cite{silvaBigData2018,SilvaASONAM2019}. Table~\ref{tab:stats_networks} shows general statistics of the networks. Overall, they present distinct characteristics that allow us to contrast the effect of our classification method on each scenario.

In summary, the academic scenario consists of 24 co-authorship networks derived from the ACM Special Interest Groups\footnote{ACM SIGs: http://www.acm.org/sigs}, as well as the full network comprising all groups with more than 79 thousand nodes and 263 thousand multiple edges. In each network, we modeled authors as nodes, coauthorships in each paper as relationships, publications' year as time intervals and tokens taken from the publication titles as attributes~\citep{silva2020knowledge}. Note that such networks are well-known in the Computer Science community, which enables us to carry out a more accurate discussion of their behavioral dynamics.

Regarding the Q\&A communities, we use the database built from the Stack Exchange network~\citep{SilvaASONAM2019}. This dataset consists of 173 Q\&A communities divided into six categories (Technology, Culture/Recreation, Life/Arts, Science, Professional and Business)\footnote{Stack Exchange: https://stackexchange.com.}. Specifically, nodes as representing community members and edges as representing answers to questions, comments to questions and comments to answers as described by Paranjape et al.~\cite{paranjape2017motifs}. In addition, each time interval has been configured to last one minute, and tokens taken from the questions and answers as attributes. The extraction process removes stop-words and reduces inflected words to their roots.

\vspace{-1mm}
\subsection{Academic Coauthorship Networks}\label{sec:analysis_academic}

\begin{figure*}[t]
    \centering
    \begin{subfigure}[b]{0.495\textwidth}
        \includegraphics[width=\textwidth]{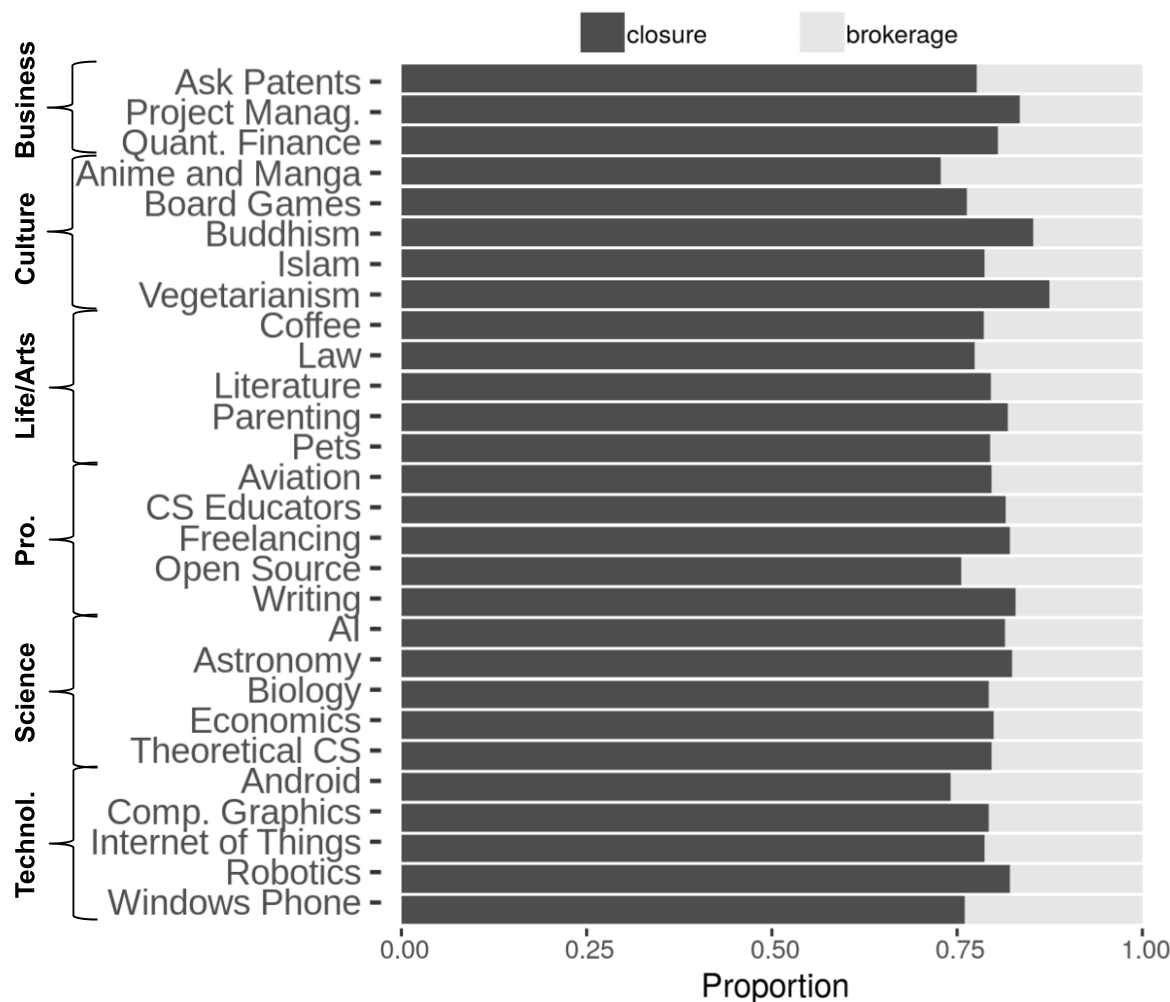}
        \caption{Nodes}
        \label{fig:classNodesQA}
    \end{subfigure}
    \begin{subfigure}[b]{0.495\textwidth}
        \includegraphics[width=\textwidth]{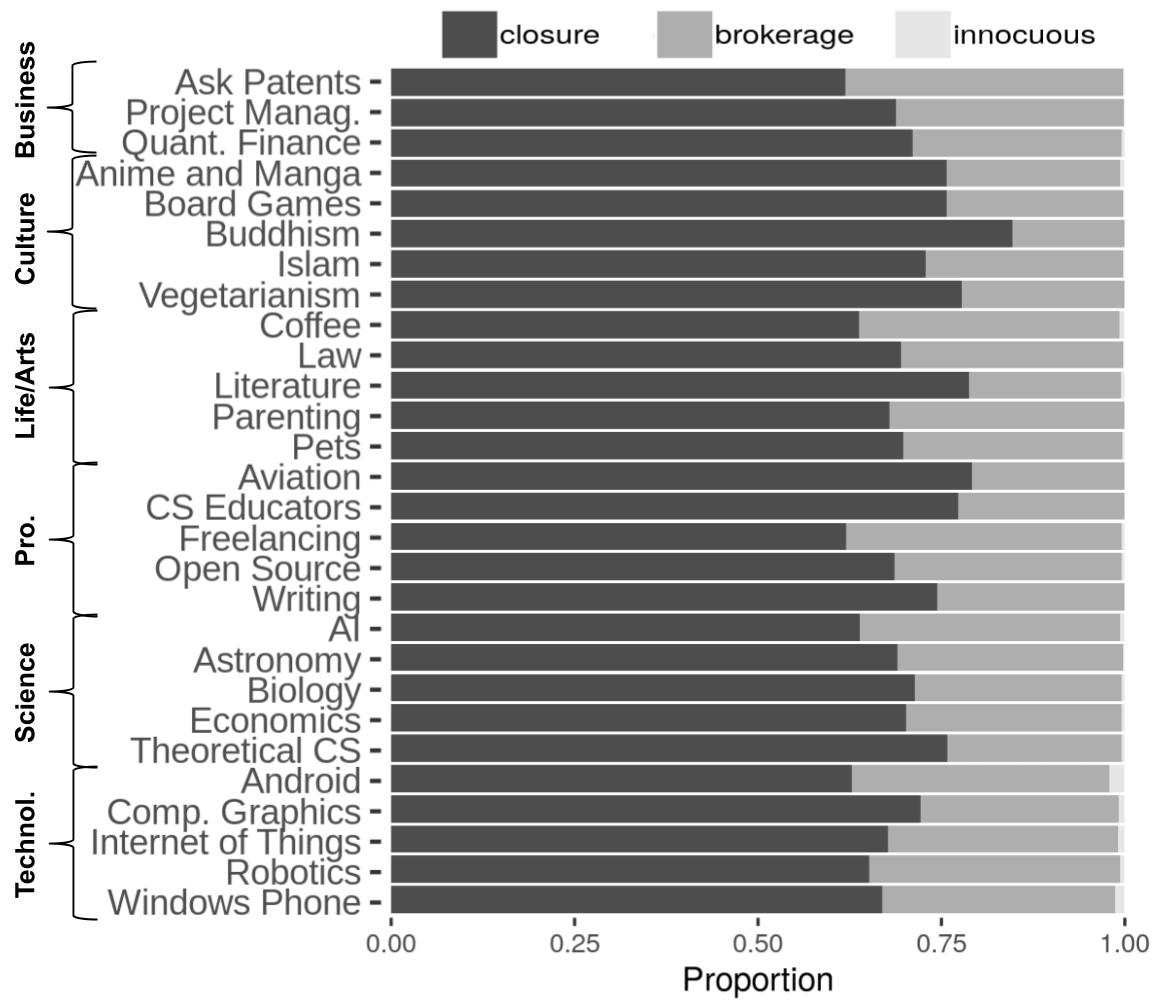}
        \caption{Edges}
        \label{fig:classEdgesQA}
    \end{subfigure}
    \caption{Node and Edge classifications for the 28 Q\&A communities.}
\end{figure*}

The \textit{Full Network} shows a substantial drop from 27.0\% to 18.8\% in the number of \textit{closure} nodes when compared with the average of all 24 ACM SIG communities. Despite that, the number of \textit{innocuous} nodes considerably increased from 57.8\% to 69.0\%. This was expected due to the fact that more active nodes (researchers) tend to participate in more than one community. Thus, with more subjects covered, the likelihood of having many relevant attributes decreases. Even so, although there are fewer \textit{closure} nodes, there is still relevant information flowing through them (i.e., high figures for \textit{closure} and \textit{brokerage} edges). In fact, such figures indicate that about 30\% of all nodes have acted structurally to make the network stronger by means of cohesion (i.e., \textit{closure}) and bridges (i.e., \textit{brokerage}).

Note also that there are small variations between the means and medians for the 24 ACM SIG communities, but with a marked standard deviation. This means that there are different social behaviors according to the specificities of each community. Specifically, Figure~\ref{fig:classNodesAcademic} presents the distribution of the node classes for the 24 ACM SIG communities and the \textit{Full Network} that includes all these communities. Overall, the classification shows a significant presence of nodes of the class \textit{innocuous} (average of 57.8\%). Indeed, an academic coauthorship network usually has a strong presence of new nodes (e.g., students or sporadic collaborators). Despite that, there is also a strong presence of nodes of the  class \textit{closure} with percentages above 30\% for more established communities such as CIKM, KDD, SIGIR, SIGMOD, STOC, SIGMETRICS, ISCA, PODC, POPL and MICRO. Particularly, most members from these communities tend to be coherent in the research topics addressed throughout their academic trajectories. In contrast, communities such as SAC, SIGUCCS, SIGGRAPH and SIGDOC show percentages below 18\% for the class \textit{closure}, which represents some lack of synergy among their members. Particularly, SIGUCCS (University and College Computing Services) and SIGDOC (Design of Communication) are two communities that address very specific topics. SIGGRAPH (Computer Graphics), although a well established scientific community, covers here only its editions up to 2003, since after that year their proceedings were discontinued and replaced by special issues of the ACM Transactions on Graphics.

Generally, such percentages can be seen as evidence of the characteristics of each community. For instance, members of the STOC (Theory of Computing) community have a tendency to show more competence in specific topics related to computation theory, thus the higher number of nodes of the class \textit{closure} (41.3\%). On the other hand, SAC (Applied Computing) is a community mainly focused on applied issues, thus covering a wide range of topics, which justifies the high number of \textit{innocuous} nodes (69.9\%).

Regarding the edge classification, Figure~\ref{fig:classEdgesAcademic} presents the distribution of the edge classes for the 24 ACM SIG networks and the \textit{Full Network}, which comprises the 24 SIG networks altogether. As we can see, most of these edge classes carry some kind of information and have been characterized as \textit{closure} or \textit{brokerage} (on average, they sum 64.7\%), thus demonstrating a strong social tie between the researchers and their relevant topics. On the other hand, edges without any social meaning (i.e., \textit{innocuous}) tend to be less present in these networks. Again, specific communities show a singular behavior, such as ISSAC (Symbolic and Algebraic Computation) and SIGIR (Research and Development in Information Retrieval) with the highest presence of \textit{closure} edges. SAC, SIGUCCS and SIGDOC also stand out for having an expressive number of \textit{innocuous} edges (more than 50\%), thus reinforcing the fact their members show no regularity with their research topics.

\subsection{Questions \& Answers Communities}\label{sec:analysis_QaA}

As we only consider frequent users in the Q\&A communities (see Section~\ref{sec:datasets}), by definition there are no \textit{innocuous} nodes in these networks~\cite{silvaBigData2018}. With respect to the node classes, Figure~\ref{fig:classNodesQA} shows few variations in the percentages of \textit{closure} and \textit{brokerage} nodes across the communities (average values of 79.8\% and 20.2\%, respectively). More specifically, the \textit{Vegetarianism} and \textit{Buddhism} communities show the highest proportions for the \textit{closure} class (87.5\% and 85.3\%, respectively), whereas \textit{Anime} \& \textit{Manga} stands for 72.8\%.

In contrast, we notice that the full academic coauthorship network had 18.8\% of its nodes  classified as \textit{closure}, 12.2\%  as \textit{brokerage} and 69.0\% as \textit{innocuous} (see Table~\ref{tab:ClassPerc}). Indeed, there are few \textit{closure} nodes (e.g., research leaders) in an academic network compared with the other ones (e.g., new students). However, in the Q\&A communities, users are in general experts and enthusiasts about specific topics, which gives them some authority~\cite{posnett2012mining,
Tang:2012, liao2018attributed, Shah2016, vasilescu2014social}.

Considering the social classification of the edges in Figure~\ref{fig:classEdgesQA}, the proportions by category and by community have significant oscillations, thus reinforcing a distinct behavior of our classification method on several topics. For example, the \textit{Buddhism} community (85.3\% of \textit{closure} nodes) has 84.7\% of \textit{closure} edges, whereas the \textit{AI} community (81.4\% of \textit{closure} nodes) has a much smaller proportion of edges (63.8\%) belonging to that same class. There are also notorious divergences between communities in the same category such as \textit{Ask Patents} and \textit{Quantitative Finance} from the \textit{Business} category, \textit{Aviation} and \textit{Freelancing} from the \textit{Professional} category, and \textit{Literature} and \textit{Parenting} from the \textit{Life/Arts} category. As we only selected frequent users, it justifies the very low presence of \textit{innocuous} edges.

By comparing the Q\&A distribution by community with the same figures from the academic ones (see Table~\ref{tab:ClassPerc}), we observed that the entire DBLP academic coauthorship network had 38.5\% of its edges classified as \textit{closure}, 41.6\% as \textit{brokerage} and 20.2\% as \textit{innocuous}. That is, we note that both scenarios reveal very different proportions of assigned classes, particularly with a higher proportion of the \textit{closure} class in the Q\&A scenarios, whereas in the academic scenarios the most representative class tended to be the \textit{brokerage} one.

\section{Evaluation of Nodes and Edges}\label{sec:experiments}

\begin{figure*}[!t]
    \centering
    \begin{subfigure}[b]{0.325\textwidth}
        \includegraphics[width=\textwidth]{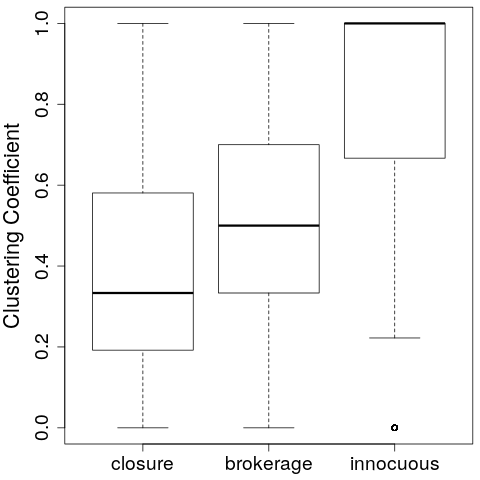}
        \caption{Clustering Coefficient}
        \label{fig:NodesCC}
    \end{subfigure}
    \begin{subfigure}[b]{0.325\textwidth}
        \includegraphics[width=\textwidth]{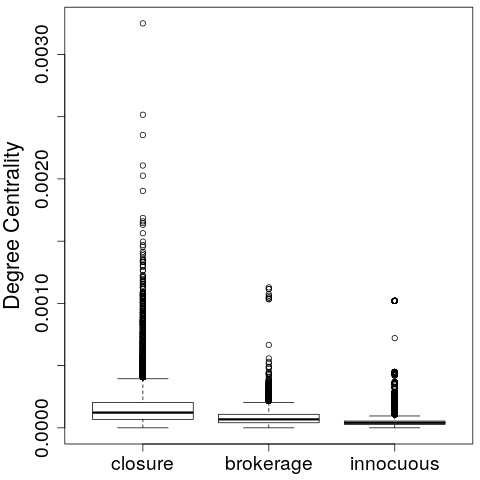}
        \caption{Degree Centrality}
        \label{fig:NodesDC}
    \end{subfigure}
    \begin{subfigure}[b]{0.325\textwidth}
        \includegraphics[width=\textwidth]{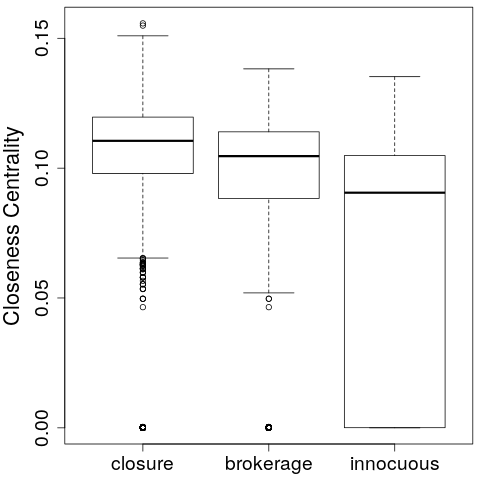}
        \caption{Closeness Centrality}
        \label{fig:NodesCL}
    \end{subfigure} \\
    \begin{subfigure}[b]{0.325\textwidth}
        \includegraphics[width=\textwidth]{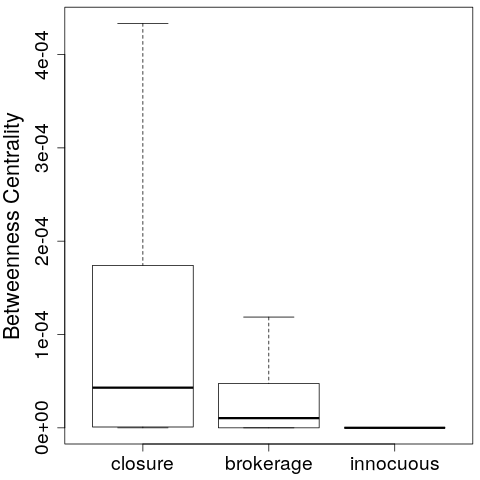}
        \caption{Betweenness Centrality}
        \label{fig:NodesBC}
    \end{subfigure}
    \begin{subfigure}[b]{0.325\textwidth}
        \includegraphics[width=\textwidth]{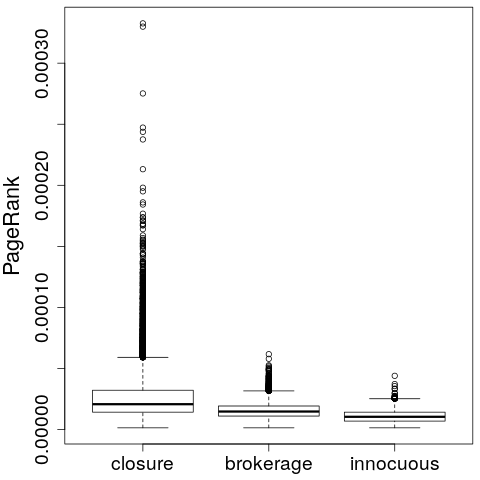}
        \caption{PageRank}
        \label{fig:NodesPR}
    \end{subfigure}
    \caption{Distribution of network properties by node class. Outliers were suppressed from graph (e) for better visualization.}\label{fig:NodesNetwork}
\end{figure*}

Given the challenging task of analyzing social interactions in a social network in order to better characterize the social role of its nodes and the meaning of its edges, Newman~\cite{Newman2004} applies network properties to evaluate the robustness of his proposed method. More specifically, he applied centrality metrics to determine the best-placed nodes in a network.

As we are also  dealing with non-labeled data, we evaluate our unsupervised classification by means of the nodes and edges acting as structurally important in a network. In others words, our methodology consists of quantifying how well nodes and edges are positioned in a social structure. For this, we explore network metrics as formally presented by Table~\ref{tab:networkMetrics}, which their social aspects are discussed as follows:

\begin{table}[!t]
\footnotesize
\centering
\caption{Network Metrics.}\label{tab:networkMetrics}
\begin{tabular}{ll}
    Metric  & Formula \\ \midrule\midrule

\makecell[l]{Degree centrality\\ of a node $i$}
 &   
$\begin{aligned}
d_i &= \frac{\sum_{j \in V} a_{ij}}{\operatorname*{arg\,max}\limits_{x \in V}~\sum_{y \in V} a_{xy}},
\end{aligned}$ \\
 &  where $a_{ij} = \left \{ \begin{array}{ll}
        1, & \text{if } (i,j) \in V  \\
        0, & \text{otherwise}
        \end{array} \right.$\\ \midrule

\makecell[l]{Closeness centrality\\ of a node $i$} &   
$\begin{aligned}
cl_i &= \frac{|V|-1}{\sum_{j \in V} d(i, j)},
\end{aligned}$
\\
 & where $d(i, j)$ is the distance between\\
    & nodes $i$ and $j$ \\ \midrule

\makecell[l]{Betweenness centrality\\ of an edge $e$} & 
$\begin{aligned}
bc_e &= \sum_{s,t \in V: s \ne t}\frac{\sigma_{st}(e)}{\sigma_{st}}
\end{aligned}$
\\
 & where $\sigma_{st}(e)$ is the number of \\
    & shortest paths from node $s$ to node $t$ \\
    & that pass through the edge $e$\\ \midrule

\makecell[l]{Betweenness centrality\\ of a node $i$} &  
$\begin{aligned}
bc_i &= \sum_{s,t \in V: s \ne t}\frac{\sigma_{st}(i)}{\sigma_{st}}, s \ne i, t \ne i
\end{aligned}$
\\
 & where $\sigma_{st}$ is the total number of\\
    & shortest paths from node $s$ to node $t$\\
    & and $\sigma_{st}(i)$ is the number of those\\
    & paths that pass through the node $i$\\ \midrule

\makecell[l]{Clustering coefficient\\ of a node $i$} &   
$\begin{aligned}
cc_i &= \frac{e_i}{n_i(n_i-1)},
\end{aligned}$
\\
 & where $e_i$ is the number of edges\\
    & between neighbors of $i$ and $n_i$ is the\\
    & number of neighbors of node $i$ \\ \midrule\midrule

\end{tabular}
\end{table}

\begin{itemize}
    \item \textit{Degree Centrality.} As 
    shown by Srinivas and Velusamy~\cite{srinivas2015identification}, this metric indicates influential nodes as, for example, a node with an immediate risk of catching a virus or getting some information. Thus, a node with high connectivity is more likely to have early access to knowledge.

    \item \textit{Closeness Centrality.} Nodes with higher closeness are, by definition, closer (on average) to the other nodes in the network. Then, we expect important classes (\textit{closure} and \textit{brokerage}) to have high values for this metric, since they have better access to knowledge from other nodes (e.g., making an opinion to reach other nodes more quickly).

    \item \textit{Betweenness Centrality.} Following Newman~\cite{Newman2004}, nodes with a high degree of betweenness centrality are likely to be influential, since they act as an intermediary for other nodes (e.g., in message-passing scenarios). Thus, as nodes and edges with high betweenness centrality values play crucial roles in the spread of knowledge in social networks~\cite{MahyarSNAM18}, then we expect high values for this metric for important nodes and edges assigned to the \textit{closure} and \textit{brokerage} classes.

    \item \textit{Clustering Coefficient.} As this metric reveals the fraction of a node's neighbors that are connected to each other (i.e., how complete the neighborhood of a node is)~\cite{srinivas2015identification}, we expect low clustering coefficient values for the most important classes (\textit{closure} and \textit{brokerage}), confirming the behavior of connecting different parts of a network.
\end{itemize}

In addition, we also use the PageRank algorithm~\cite{page1999pagerank} by  considering that more important nodes tend to make stronger endorsements due to their connectivity and ties to other important nodes. That is, we also expect \textit{closure} and \textit{brokerage} nodes to have high values for this metric.

\subsection{Results}

By means of the aforementioned social properties, we now assess the importance of nodes and their dynamics relationships by contrasting the classes assigned to them with the network properties\footnote{All experiments were performed with a significance level of $\alpha = 0.05$.}. 
Considering the node classification, Figure~\ref{fig:NodesNetwork} shows the distribution of the network properties by node class. In Figure~\ref{fig:NodesCL}, we can see that the lowest clustering coefficient values are for nodes classified as \textit{closure}, followed by those classified as \textit{brokerage}. On the other hand, nodes classified as \textit{innocuous} tend to have worse positions in the social structure. In this way, these results indicate that nodes classified as \textit{innocuous} are highly dependent on their neighborhood, while those classified as \textit{closure} and \textit{brokerage} tend to diversify their relationships. In addition, Figure~\ref{fig:NodesDC}~and~\ref{fig:NodesCC} confirm that nodes classified as \textit{closure} and \textit{brokerage} tend to have a better position in the social structure as, respectively, having more connections in the network (high degree centrality) and being on average closer to other nodes (high closeness centrality). 

Considering the importance of nodes in terms of the paths that pass through them and how much endorsement they receive, Figures~\ref{fig:NodesBC}~and~\ref{fig:NodesPR} respectively show that nodes classified as \textit{closure} and \textit{brokerage} have more information passing through them and are seen as topologically more relevant by their peers.
Note that all cases have a clear class distinction. Formally, all distributions are statistically different by means of the Mann-Whitney-Wilcoxon test (among all classes) and by the Kruskal-Wallis test (between each pair)~\cite{hollander2013nonparametric}.

\begin{figure*}[!t]
    \centering
    \begin{subfigure}[b]{0.325\textwidth}
        \includegraphics[width=\textwidth]{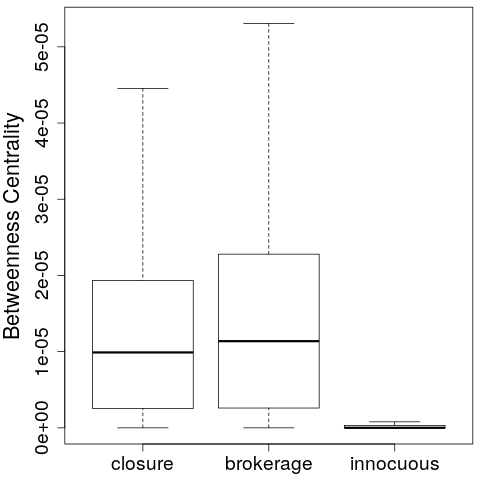}
        \caption{Our method}
        \label{fig:EdgesBC}
    \end{subfigure}
    ~~
    \begin{subfigure}[b]{0.325\textwidth}
    \includegraphics[width=\textwidth]{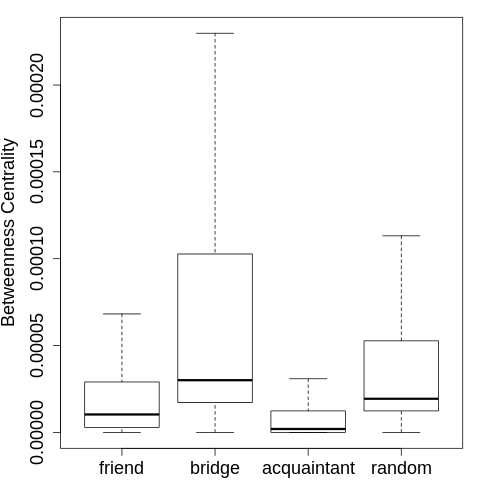}
    \caption{RECAST}
    \label{fig:EdgesRecast}
    \end{subfigure}
    \caption{Distribution of betweenness centrality values by edge class. Outliers were suppressed for better visualization.}\label{fig:EdgesNetwork}
\end{figure*}

As for the classification of the edges, Figure~\ref{fig:EdgesBC} shows the distribution of the betweenness centrality metric with respect to our classification. We clearly note that the \textit{brokerage} and \textit{closure} classes have more expressive values for this metric. Note that the distributions of \textit{closure} and \textit{brokerage} distinguish less than those reported for nodes, but now the \textit{brokerage} class is slightly superior to the \textit{closure} one in contrast to the classification of the nodes (Figures~\ref{fig:NodesNetwork}a-e). Nonetheless, they are still statistically different according to the Kruskal-Wallis and Mann-Whitney-Wilcoxon tests. Moreover, even though the \textit{innocuous} class accounts for 31.4\% of all edges, their centrality values are very low.

For the sake of comparison, Figure~\ref{fig:EdgesRecast} depicts the same distribution for the RECAST classes. As discussed in Section~\ref{sec:related}, the RECAST algorithm assigns social classes to edges in temporal networks. For this, it explores the regularity of the relationships and the topological overlap existing among them over time. By comparing such regularities with random temporal graphs, it classifies social ties as \textit{friend}, \textit{bridge}, \textit{acquaintant} and \textit{random}. In this way, we expect the important social classes (\textit{friend} and \textit{bridge}) to have better network properties than those considered less important (\textit{acquaintant} and \textit{random}).

Similar to our classification, the most expressive values of centrality are those assigned to the \textit{bridge} class. On the other hand, RECAST classifies many structural edges (i.e., those with high network properties) as \textit{random}, as well as several edges with low figures as \textit{friend}. In conclusion, our method brings a new perspective and provides a more accurate analysis to characterize such relationships. 

\vspace{-1mm}
\subsection{Sensitivity Analysis}

Since we are dealing with temporal attributed networks, the relevant attributes and time aspects must be properly analyzed regarding the effectiveness of our classification method. Accordingly, we address these issues next. \\

\noindent
\textit{Discriminatory power of the attributes.} In order to measure the strength of social interactions, Algorithm~\ref{algo:extrairAtributos} ensures the function $\Gamma$ containing the sets of all statistically relevant attributes for each node. In fact, if an attribute is associated with a node several times, then we can infer its importance.

However, a specific statistical treatment can be added to this process in order to exclude attributes that, even if randomly distributed, were erroneously considered as relevant ones. This additional statistical step consists in making the function $\Phi$, which associates each edge $e$ with a specific set of attributes, a random association $\Phi'$. Then, we get $\Gamma$ from different $\Phi'$ instances to measure the probability that each attribute has been erroneously classified as being \textit{relevant}. Finally, we exclude such attributes that were considered as relevant with probability significantly higher than the level of significance $\alpha$. In other words, we filter from our input the attributes that can interfere in the process of identifying the relevant ones. Even removing some of the data, we expect the proposed method to be robust enough to properly classify nodes and edges.

As a result, both configurations (without the exclusion step and with the step of excluding attributes that are not statistically valid when randomly distributed) are statistically equivalent by means of the distribution of network properties by classes. \\

\noindent
\textit{Existence time of the nodes.} This sensitivity test consists in investigating the robustness of our approach to differentiate nodes with similar existence times. For this, we divided the nodes into the following annual time intervals: $[1,5)$, $[5,10)$, $[10,15)$ e $[15, \infty)$.
Our method was able to distinguish the distributions of all network metrics by classes for all time intervals in terms of the Kruskal-Wallis test. However, for the time interval $[1,5)$, the Mann-Whitney-Wilcoxon test did not differentiate the distributions between the classes \textit{closure} and \textit{brokerage} for the metrics \textit{betweenness centrality} and \textit{clustering coefficient}.

\section{Conclusions and Future Work}\label{sec:conclusions}

In this article, we reinforce the importance of the network theory paradigm for understanding the complexity that involves real world actors and their relationships~\cite{barabasi2009scale}. Based on the \textit{structural autonomy} that captures when people are tightly connected to one another with extensive bridge ties beyond them~\cite{burt2005brokerage}, we emphasize the concept of \textit{closure} as representing the importance of a node in terms of its expertise according their associated attributes (strong ties), whereas the \textit{brokerage} one captures its potential for transferring its attributes (weak ties). Then, we proposed a node-attribute graph model that captures the social tie of individuals and their associated attributes, thus exploring the importance of the persistence of node-attribute relationships over time.

Overall, our classification method was able to reveal the social role of the nodes and the strength of the social meaning of their multiple interactions in different social contexts from academic coauthorship networks and Q\&A communities. For instance, there is a contrasting social behavior when comparing the \textit{Theory of Computing} and \textit{Applied Computing} networks, as well as when we compare the \textit{Buddhism} and \textit{Islam} communities. In addition, based on Newman's experimental methodology~\cite{Newman2004}, we statistically validated the assigned classes according to network properties, thus agreeing with their expected social meaning.

As future work, we aim to apply our social-based characterization approach to the problem of community detection \cite{LeaoJISA2018,dao_bothorel_lenca_2020}. We also intend to propose a new strategy to explore its propagation behavior, mainly focusing on knowledge transfer aspects (i.e., characterizing how the attributes pass through the network)~\cite{silva2020knowledge}. In addition, we plan to investigate the persistence of the nodes with respect to their neighborhood in order to identify the most influential ones.

\section*{Acknowledgments}
\noindent
This work was supported by the authors' individual grants from CAPES and CNPq.

\bibliographystyle{ACM-Reference-Format}
\bibliography{paper}

\end{document}